\documentclass{aa}  

\usepackage{graphicx}
\usepackage{txfonts}
\usepackage[T1]{fontenc}
\usepackage{ae,aecompl}
\pdfoutput=1

\usepackage{lscape}
\usepackage{color}
\usepackage{tabularx}

\usepackage{graphicx}   
\usepackage{amsmath}    
\usepackage{amssymb}    

\begin{document} 

\title{Accurate stellar parameters and distance to two evolved eclipsing binary systems, OGLE-BLG-ECL-123903 and OGLE-BLG-ECL-296596, towards the Galactic bulge.}
\titlerunning{Accurate stellar parameters of two evolved eclipsing binary systems.}
\author{K. Suchomska\inst{1}\thanks{E-mail: ksenia@astrouw.edu.pl (KS)}
\and D. Graczyk\inst{2,3,5}
\and G. Pietrzy{\'n}ski\inst{2}
\and W. Gieren\inst{3,5}
\and J. Ostrowski \inst{4}
\and R. Smolec \inst{2}
\and A. Tkachenko\inst{6}
\and M. G{\'o}rski\inst{3,5}
 \and P. Karczmarek\inst{1}
 \and P. Wielg{\'o}rski\inst{2}
\and B. Zgirski\inst{2}
\and I. B. Thompson\inst{7}
\and S. Villanova\inst{3}
\and B. Pilecki \inst{2}
\and M. Taormina\inst{2}
\and Z. Ko{\l}aczkowski\inst{2}
  \and W. Narloch\inst{2}
  \and I. Soszy{\'n}ski \inst{1}}

\institute{Warsaw University Observatory, Al. Ujazdowskie 4, 00-478 Warsaw, Poland
\and Nicolaus Copernicus Astronomical Centre, Bartycka 18, 00-716 Warsaw, Poland
\and Departamento de Astronom{\'i}a, Universidad de Concepci{\'o}n, Casilla 160-C, Concepci{\'o}n, Chile
\and Pedagogical University of Cracow, ul. Podchor\k{a}\.{z}ych 2, 30-084, Krak\'{o}w, Poland
\and Millenium Institute of Astrophysics, Av. Vicu{\~n}a Mackenna 4860, Santiago, Chile
\and Instituut voor Sterrenkunde, KU Leuven, Celestijnenlaan 200D, B-3001 Leuven, Belgium
\and Carnegie Observatories, 813 Santa Barbara Street, Pasadena, CA 911101-1292, USA}

\date{Received date/ accepted date}




\abstract{}{Our aim is to obtain high-accuracy measurements of the physical and orbital parameters of two evolved eclipsing binary systems, and to use these measurements to  study their evolutionary status. We also aim to derive the distances to the systems by using a surface brightness--colour relation and compare these distances with  the measurements provided by GAIA.}  
{We measured the physical and orbital parameters of both systems based on $V$-band and $I$-band photometry from OGLE, near-infrared photometry obtained with the NTT telescope and the instrument SOFI, as well as high-resolution spectra obtained at ESO 3.6m/HARPS and Clay 6.5/MIKE spectrographs. The light curves and radial-velocity curves were analysed with the Wilson--Devinney code. }
{We analysed two double-lined eclipsing binary systems OGLE-BLG-ECL-123903 and OGLE-BLG-ECL-296596 from the Optical Gravitational Lensing Experiment (OGLE) catalogue. Both systems have a configuration of two well-detached giants stars. The masses of the components of OGLE-BLG-ECL-123903 are $M_1$ = 2.045 $\pm$ 0.027 and $M_2$ = 2.074 $\pm$ 0.023 $M_\odot$ and the radii are  $R_1$ = 9.540 $\pm$ 0.049 and $R_2$ = 9.052 $\pm$ 0.060 R$_\odot$. For OGLE-BLG-ECL-296596, the masses are $M_1$= 1.093 $\pm$ 0.015 and $M_2$= 1.125 $\pm$ 0.014 $M_\odot$, while the radii are $R_1$= 18.06 $\pm$ 0.28 and $R_2$= 29.80 $\pm$ 0.33 $R_\odot$. Evolutionary status was discussed based on the isochrones and evolutionary tracks from \textsc{parsec} and \textsc{mesa} codes. The ages of the systems were established to be around 1.3 Gyr for the OGLE-BLG-ECL-123903 and 7.7 Gyr for the OGLE-BLG-ECL-296596. We also determined the distance to both systems. For OGLE-BLG-ECL-123903 this is equal to $d$= 2.95 $\pm$ 0.06 (stat.) $\pm$ 0.07 (syst.) kpc, while for the OGLE-BLG-ECL-296596 it is $d$= 5.68 $\pm$ 0.07 (stat.) $\pm$ 0.14 (syst.) kpc. This is the first analysis of its kind for these unique evolved eclipsing binary systems.}{}

\maketitle


\section{Introduction}

One of the most important tasks in modern astronomy is to understand the basic physics and evolution of stars. Despite great progress in the theory of evolution and stellar structure due to the high level of theoretical models of star evolution, the process itself is not still fully understood. The comparison of more and more precise observations with those models suggests some inaccuracies, indicating that some aspects, such as treatment of convection, should be refined. 
Stellar-evolution models rely on accurate determinations of stellar parameters such as radius, mass, and effective temperature. \cite{torres10} noted that only parameters derived with an accuracy of $\sim$ 1-3$\%$ provide sufficiently strong constraints for models with inadequate physics to
be rejected.  Well-detached eclipsing binary systems consisting of two evolved stars are great candidates for such analyses. Calculations of their orbits allow us to directly determine the masses of their components with the required accuracy, which also gives us a chance to estimate other physical parameters. 
It is worth mentioning that eclipsing binary systems, where the components are evolved stars, are rarely found, due to their long orbital period and very narrow eclipses. The Araucaria Project, based on the analysis of the photometric data for over 50 millions stars in the centre of the Milky Way, as well as in the Magellanic Clouds, collected by the OGLE project, selected over a dozen very unique detached eclipsing binary systems with two giants stars.

We present photometric and spectroscopic observations  of two double-lined eclipsing binary systems from the Optical Gravitational Lensing Experiment catalogue (OGLE) identified as OGLE-BLG-ECL-123903 and OGLE-BLG-ECL-296596 (hereafter BLG-123903 and BLG-296596, respectively) \citep{Soszynski2016}. 
Both binaries consist of two evolved giant stars and are located towards the Galactic bulge.
Basic data of the systems are presented in Table~\ref{tab:basic}. Few such systems  \citep[e.g.][]{hel15, suchomska15} in our Galaxy have been studied in detail. 
In this paper we present a precise determination of the physical and orbital parameters, together with the derived distances to the binaries. 
Section 2 presents details of the photometric and spectroscopic data, Section 3 outlines the analysis of the data and the modelling of the two systems, Section 4 presents an analysis of the evolutionary status and distance to these two binaries, and Section 5 presents a summary of our results.

\begin{table} 
\caption{Basic data for OGLE-BLG-ECL-123903 and OGLE-BLG-ECL-296596.}
\centering
  \resizebox{\columnwidth}{!}{\begin{tabular}{ccc}
   \hline
  & BLG-ECL-123903 & BLG-ECL-296596\\ 
  \hline
  2MASS & J17494566-2918071 & J18052042-2625530\\
  GAIA DR2 & 4057237012210464512  & 4063342909579750528  \\
  $\alpha$&17:49:45.68 & 18:05:20.43\\
  $\delta$&-29:18:07.2 &-26:25:53.2 \\
  Period (d)& 118.53& 155.56\\
  $V$ (mag) $^1$&16.904 $\pm$0.009& 15.483 $\pm$ 0.005\\
  $I$ (mag) $^1$& 14.083 $\pm$ 0.003& 13.074 $\pm$ 0.007\\
  $J$ (mag) $^2$& 11.952 $\pm$ 0.031& 11.297 $\pm$ 0.062 \\
  $K$ (mag) $^2$& 10.824 $\pm$ 0.053& 10.146 $\pm$ 0.053 \\
  $J$ (mag) $^3$&12.016 $\pm$ 0.025&11.282 $\pm$ 0.028\\
  $K$ (mag) $^3$& 10.746 $\pm$ 0.047& 8.913 $\pm$ 0.027\\ \hline
  \multicolumn{2}{l} {$^1$ -OGLE-IV Johnson-Cousins filters} \\
  $^2$ - UKIRT \\
  $^3$ - 2MASS \\
   \end{tabular}}
  \centering
  \label{tab:basic}
\end{table}

\section{Observations}

\subsection{Photometry}
In our analysis we used the $V$-band and $I$-band optical photometry obtained with the Warsaw 1.3 m telescope at Las Campanas Observatory during the third and fourth phase of the OGLE project \citep{Udalski2003,Soszynski2012,Soszynski2016}. 
For BLG-123903 a total of 110 and 4326 measurements were obtained in the $V$-band and $I$-band filters, respectively. The data coverage for the light curve for this system is complete in both filters. For the analysis of BLG-296596 a total of 785 measurements were obtained in the $I$-band  and the light curve coverage was complete. In the $V$-band  only 13 measurements were taken, from which  we used 7 measurements obtained outside the eclipses in order to determine the luminosity of this system in the $V$-band. 

We also collected near-infrared  $J$-band and $K$-band photometry with the use of the ESO NTT telescope at La Silla Observatory equipped with the SOFI camera. We obtained six epochs of infrared photometry for each system, all taken outside of the eclipses. The setup of the instrument, as well as the process of reduction and calibration of the  data onto the UKIRT system are described in \cite{pietrzyn2009}. The infrared photometry was then transformed onto the 2MASS system using the prescription presented in \cite{car01}.

\subsection{Spectroscopy}
For our analysis we used high-resolution spectra that we collected with the Clay 6.5-m telescope at Las Campanas Observatory, equipped with the MIKE spectrograph, as well as spectra collected with the ESO 3.6-m telescope at La Silla Observatory, equipped with the HARPS spectrograph. For the  MIKE spectrograph observations we used the 5 x 0.7 arcsec slit which gave a resolution of $\sim$42 000. The  HARPS spectrograph was used in the EGGS mode at a resolution of $\sim$80 000. 

For our analysis of the BLG-123903 system we used 15 spectra in total, 8 of which were taken with the MIKE spectrograph and 7 with the HARPS spectrograph. For the BLG-296596 analysis  we collected 21 spectra in total, 11  with the MIKE spectrograph and 10 with the HARPS spectrograph. 

\section{Analysis and Results}
\label{analysis}
Absolute physical and orbital parameters of our systems were derived using the Wilson--Devinney code (hereafter WD code), version 2007  \citep{wil07, wilson71, wilson79, wilson90}, equipped with the automated differential correction (DC) optimizing subroutine. The WD code simultaneously solves the multiband light curves and radial-velocity curves, which is essential to obtain a consistent model of the binary system. 

We also used the JKTEBOB code \citep{south2004,south2005}, which includes a Monte Carlo error analysis algorithm in order to compare the errors of the derived parameters obtained directly from the WD code with those from the Monte Carlo simulations.  

To measure radial velocities of the components of the analysed systems we used the RaveSpan software  \citep{pilecki2013, pilecki2015, pilecki2017} which uses the Broadening Function formalism \citep{ruc92, ruc99}. Template spectra were selected from the synthetic library of LTE spectra  of \citet{coelho05}. This code also allowed us to estimate the spectroscopic luminosity ratios of the components which were  compared to the luminosity ratios from the WD solution to check for consistency in the results. This software was also used to perform spectrum disentangling. An example of an application of the RaveSpan code can be found in \cite{pilecki2018}.

\subsubsection{BLG-123903}

Radial velocities of the components of the BLG-123903 system were measured over the wavelength range of 5160--6800 \AA. We created a mask that excludes H~$\alpha$ as well as telluric lines. We also applied the velocity offset between HARPS and MIKE data, correcting the radial velocity measurements of MIKE spectra by -0.29 km  s$^{-1}$. Our measurements of the radial velocities of both components are presented in Tab.~\ref{tab:velocities1}.

In order to check if the components rotate synchronously, we measured their rotational velocities and compared them with the expected synchronous velocities. 
To determine the rotational velocities we used RaveSpan code fitting rotationally broadened profiles. For our measurements we used seven spectra obtained with the MIKE spectrograph. The mean values of the  broadenings of the primary and secondary components were measured to be $v_{M_1}$ = 9.99$\pm$ 0.51 km s$^{-1}$ and $v_{M_2}$= 9.99 $\pm$ 0.44 km s$^{-1}$. 

In order to determine the correct $v$sin$i$ we also took into account the macroturbulence velocities and the instrumental profile contributions, as described in \cite{suchomska15}.
The macroturbulance velocities were estimated to be $v_{mt1}$=1.74 km s$^{-1}$ and $v_{mt2}$=1.70 km s$^{-1}$  for the primary and secondary components. Assuming the resolution of the MIKE spectrograph to be $R$=42 000, we estimated the instrumental profile to be $v_{ip}$= 4.28 km s$^{-1}$. The resulting rotational velocities are  $v_1\sin{i}$ = 8.86 $\pm$ 1.51 and $v_2\sin{i}$ = 8.86 $\pm$ 1.44 km s$^{-1}$.

The expected equatorial rotational velocities are $v_1$ = 4.07 km s$^{-1}$ and $v_2$ = 3.87 km s$^{-1}$. 
The measured rotational velocities are not consistent with the expected values, and we therefore conclude that the components do not rotate synchronously. 
We also compared our measured $v$sin$i$ with the pseudo-synchronous radial velocities. \cite{Hut1981} gives the definition of the pseudo-synchronization, describing the phenomenon of near synchronization of revolution and rotation around periastron,
where the tidal interaction is the strongest. In order to obtain pseudo-synchronous radial velocities, we multiplied the expected rotational velocities by  $\sqrt{(1+e)/(1-e)^3}$. The expected pseudo-synchronous velocities are $v_{ps1}$= 8.12 km s$^{-1}$ and $v_{ps2}$= 7.72 km s$^{-1}$. Comparing these values with the measured ones, we conclude that the components rotate pseudo-synchronously.

\subsubsection{BLG-296596}

Radial velocities of the components of the BLG-296596 system were calculated over the wavelength range of 5350--6800 \AA. We also applied the velocity offset between HARPS and MIKE data, correcting the radial velocity measurements of MIKE spectra by -0.10 km  s$^{-1}$. The measurements of the radial velocities of both components are presented in Table~\ref{tab:velocities2}. We noticed that the components differ in the systemic velocity, and therefore applied a correction of $v_{corr}$= -703 m s$^{-1}$ to the radial velocities of the first component during our WD modelling. The difference in the systemic velocity of the components can be caused either by large-scale convective motions or by a differential gravitational redshift between the stars (e.g. \cite{torres09}).

To check for the synchronous rotation we performed the same analysis as for  BLG-123903. We used ten spectra obtained with the MIKE spectrograph and fitted the rotationally broadened profiles with the RaveSpan code. The mean values of the measured broadenings are  $v_{M_1}$ = 9.85$\pm$ 0.29 km s$^{-1}$ and $v_{M_2}$= 19.25 $\pm$ 0.31 km s$^{-1}$ for the primary and secondary components. The macroturbulence velocities for the components were determined to be $v_{mt1}$=1.35 km s$^{-1}$ and $v_{mt2}$=1.33 km s$^{-1}$. The instrumental profile was estimated to be $v_{ip}$= 4.28 km s$^{-1}$. 
The resulting rotational velocities of the components were estimated to be $v_1\sin{i}$ = 8.76 $\pm$ 1.29 and $v_2\sin{i}$ = 18.72 $\pm$ 1.31 km s$^{-1}$. 
The expected equatorial rotational velocities were determined to be $v_1$ = 5.88 km s$^{-1}$ and $v_2$ = 9.69 km s$^{-1}$. If we compare the measured rotational velocities with the expected values we conclude that the components do not rotate synchronously. However, as for BLG-123903, we checked for pseudo-synchronous rotation. The pseudo-synchronous velocities were estimated to be $v_{ps1}$= 8.92 km s$^{-1}$ and $v_{ps2}$= 14.73 km s$^{-1}$, and we conclude that for this system only the primary component rotates pseudo-synchronously.

\begin{table} 
\caption{Radial-velocity measurements of  BLG- 123903. }
\centering
 \begin{tabular}{p{0.22\linewidth}p{0.19\linewidth}p{0.2\linewidth}p{0.19\linewidth}}
   \hline
HJD & V$_1$ & V$_2$ &Instrument\\ 
  --2450000 & (km s$^{-1}$) & (km s$^{-1}$) &\\ \hline \hline
6102.58981 & -44.450 &  -8.503 & MIKE\\
6449.77559 & -49.835 &  -2.327  & HARPS\\
6490.71411 &   -6.611  &        -45.081  & MIKE\\
6554.55147 & -54.014 &  2.619  & HARPS\\
6558.61555 & -52.656 &  3.389  & MIKE\\
6560.52555 & -52.862 &  1.448  & MIKE\\
6571.52849 & -47.483 &  -3.901  & HARPS\\
6572.51901 &  46.352 &  -4.901  & HARPS\\
6844.54902 &   -8.694 & -41.795  & MIKE\\
6878.58334 &   -8.442 & -42.013  & HARPS\\
6908.53121 & -54.264 &  3.438   & HARPS\\
6931.56154 & -44.051 &  -6.439  & MIKE\\
7160.82519 & -48.789 &  -1.595   & HARPS\\
7210.77289 &    8.779 & -58.850  & MIKE\\
7212.53500 &  11.539 &  -60.716   & MIKE\\ \hline
  \end{tabular}
  \centering
  \label{tab:velocities1}
\end{table}

\begin{table} 
\caption{Radial-velocity measurements of  BLG- 296596. }
\centering
  \begin{tabular}{p{0.22\linewidth}p{0.19\linewidth}p{0.2\linewidth}p{0.19\linewidth}}
   
   \hline
   HJD & V$_1$ & V$_2$ &Instrument\\ 
  --2450000 & (km s$^{-1}$) & (km s$^{-1}$) &\\ \hline \hline
6097.84020 &  -29.781 &    -67.465 &    MIKE\\
6102.53422 &  -34.692 &   -63.141  &   MIKE\\
6448.85378 &  -63.566 &    -35.975 &   HARPS\\
6462.68675 &  -68.934 &    -30.100 &   HARPS\\
6487.61875 &  -69.774 &    -29.361 &   MIKE\\
6490.63957 &  -69.754 &    -32.223 &   MIKE\\
6529.55037 &  -30.275 &    -68.902 &   HARPS\\
6552.51582 &  -21.427 &    -78.194 &   HARPS\\
6554.49049 &  -22.566 &    -77.282 &   HARPS\\
6558.49692 &  -24.938 &    -72.817 &    MIKE\\
6559.52407 &  -25.138 &    -72.532 &   MIKE\\
6572.59194 &  -38.264 &    -60.100 &    HARPS\\
6842.68322 &  -27.506 &    -69.856 &    MIKE\\
6845.73979 &  -24.959 &    -73.871 &   MIKE\\
6846.57158 &  -23.429 &    -74.615 &    MIKE\\
6846.72728 &  -24.222 &    -75.046 &   MIKE\\
6908.56402 &  -59.267 &    -40.446 &    HARPS\\
6909.55300 &  -59.979 &    -38.639 &   HARPS\\
6931.55082 &  -69.556 &    -30.124 &   MIKE\\
7159.76838 &  -22.539 &    -76.465 &    HARPS\\
7161.78864 &  -21.407 &    -78.160 &   HARPS\\ \hline
  \end{tabular}
  \centering
  \label{tab:velocities2}
\end{table}

\subsection{Spectral disentangling and atmospheric analysis}
\label{spec}
To disentangle the spectra we used the RaveSpan code which uses a disentangling procedure based on the method presented  by \cite{gon06}. The method works in the real wavelength domain and requires  high signal-to-noise-ratio(S/N) data. Therefore for both systems we used only the MIKE spectra. To derive properly renormalized disentangled spectra, we followed a two-step process presented in \cite{gra14}. The disentangled spectra were used for an atmospheric analysis of the individual components. We assumed local thermodynamical equilibrium and used the MOOG program \citep{sne73} to derive effective temperatures ($T_{\rm eff}$), gravities (log~$g$) and metallicities ([Fe/H]). We adopted the line list given by \cite{vil10}. The derived parameters for both systems are presented in Tables ~\ref{tab.atmo1} and ~\ref{tab.atmo2}. The derived effective temperatures were used as input to the WD models and also to estimate the interstellar extinction toward the two systems.

Given the inaccuracies associated with spectral disentangling and the uncertain metallicity determinations, we concluded that the difference in the metallicity derived for  the components in both of the systems analysed is not significant.  In our analysis with the WD code, we assumed a metallicity of [Fe/H] = 0.2 dex for  BLG-123903 and [Fe/H] = 0.0 dex for  BLG-296596.  

\begin{table}
\caption{Atmospheric parameters of the components of BLG-123903.}
\begin{tabular}{p{0.22\linewidth}p{0.19\linewidth}p{0.2\linewidth}p{0.19\linewidth}}
\hline
Component & $T_{eff}$ [K] &[Fe/H] & log $g$\\ \hline
Primary & 4860 $\pm$ 185&0.14$\pm$ 0.23 & 2.79*\\
Secondary & 4700 $\pm$ 180& 0.39 $\pm$0.28& 2.84*\\ \hline
\multicolumn{2}{l}{* - fixed from WD solution}
\end{tabular}
\centering
\label{tab.atmo1}
\end{table}

\begin{table}
\caption{Atmospheric parameters of the components of BLG-296596.}
\begin{tabular}{p{0.22\linewidth}p{0.19\linewidth}p{0.2\linewidth}p{0.19\linewidth}}
\hline
Component & $T_{eff}$ [K] &[Fe/H] & log $g$\\ \hline
Primary & 4215 $\pm$ 135&-0.02$\pm$ 0.23 & 1.96*\\
Secondary & 4050 $\pm$ 120& -0.11 $\pm$0.19& 1.54*\\ \hline
\multicolumn{2}{l}{* - fixed from WD solution}
\end{tabular}
\centering
\label{tab.atmo2}
\end{table}

\subsection{Interstellar extinction}

\subsubsection{BLG-123903}
We used   effective temperature -- $(V-K)$ colour calibrations to derive the interstellar reddening of BLG-123903.
We used several calibrations of $T_{\rm eff}$-- $(V-K)$ colour \citep{ben98,alo99,hou00,ram05,mas06, gon09,cas10,wor11}, using the values of $T_{\rm eff}$  obtained during atmospheric analysis (Tab.~\ref{tab:results_final1}). We estimated a value of $E(B-V)$= 1.329$\pm$0.114 mag. The error of the $E(B-V)$ determination is the result of the accuracy of $T_{\rm eff}$, the accuracy of the adopted effective temperature--colour calibration, the accuracy of the photometric solution for the system, and the accuracy of the initial $V-$ and $K$-band photometry.

We also used the extinction maps of \cite{schle98} with the recalibration of \cite{schla11} to estimate the reddening in the direction of BLG-123903. We followed the description given in \cite{suchomska15}. We estimated the value of  $E(B-V)$ = 0.329 $\pm$ 0.059.

We also estimated the reddening toward BLG-123903 using  results from the 3D Dust Mapping which is based on Pan-STARRS~1 photometry of 800 million stars together with 2MASS photometry of 200 million stars (\cite{green15, green18}. The authors present  3D maps of interstellar dust reddening which trace dust reddening both as a function of angular position on the sky and distance. Based on the position and distance to BLG-123903, the reddening to the system is estimated to be E(B-V) = 0.34 $\pm$ 0.05.

The reddening derived from both Schlegel maps and Pan-STARRS~1 shows a substantial disagreement with the result determined from effective temperature -- ($V-K$) colour calibrations. This result might mean that towards the direction of BLG-123903 there is a significant interstellar cloud. This shows the extent of the  heterogeneous distribution of the interstellar matter, especially in the plane of the Galactic disc  and towards the centre of the Milky Way. 

We decided to adopt the reddening estimate based on $T_{\rm eff}$-- $(V-K)$ colour relations alone,  $E(B-V)$ = 1.329 $\pm$ 0.114. 

\subsubsection{BLG-296596}

We used the same methodology to estimate the reddening to BLG-296596. The results were:
$E(B-V)$ = 0.732 $\pm$ 0.093 mag ($T_{\rm eff}$-- $(V-K)$ colour relations), $E(B-V)$ = 0.681$ \pm$ 0.058 mag
(Schlegel maps), and $E(B-V)$ = 0.88  $\pm$ 0.093 mag (3D Dust Mapping). We adopted the average
of these three values, $E(B-V)$ = 0.731 $\pm$ 0.117, as the reddening toward BLG-296596, where
the estimated error is a combination of statistical and systematic error.

\subsection{Modelling}

We used the WD code to model both systems. This code is based on Roche lobe geometry and allowed us to simultaneously fit $I$-band light curves along with the radial velocity curve to a geometric model of a detached binary system.

\subsubsection{BLG-123903}

The orbital period and the moment of primary minimum were originally adopted from the OGLE catalogue \citep{Soszynski2016}: $P$=118.529 d and $T{_0}$=2456644.2.  The orbital period was later adjusted along with the phase shift parameter (PSHIFT). The main purpose of using the PSHIFT parameter is to allow the   subroutine in the WD code to adjust for a zero point error in the ephemeris used to compute the phases. We also derived the average out-of-eclipse magnitudes using all of the observational data outside of minima. We measured  mean values of  $V$ = 16.904 and  $I$ = 14.083. In this work we refer to the primary component as the star being eclipsed during the deeper, primary minimum.  

We used the WD code to perform the modelling of the system. We simultaneously fitted    the $V-$ and $I$-band light curves and radial-velocity curves. In order to determine the best-fit model it is crucial to decide which parameters are adjusted as this has a significant impact on the final solution and the parameters of the model. We chose
the input parameters for the DC subroutine as described in \cite{gra12}. 
The choice of adjustable parameters, albedo and gravity brightening parameters, as well as the adopted limb darkening law, was as described in \cite{suchomska15}. Additionally, we adjusted the orbital eccentricity ($e$) and the argument of periastron ($\omega_0$).

   To set the temperature scale for each component we adopted an initial value of the temperature of the primary of $T_1$=5500~K. We used this as a starting point for our analysis and then iterated using the DC subroutine of the WD code in order to find a best solution for both $V-$ and $I$-band light curves, as well as the radial velocity curve. During this procedure all free parameters were adjusted at the same time. 
   
   
   At the end of the fitting procedure we also checked for the presence of the third light ($I_3$) parameter to determine its impact on the solution. However the solution suggested an unphysical value for $I_3$, and therefore we adopted $I_3$ = 0 in our final solution. 
   
   The luminosity ratios of the components obtained from the WD solution were compared with the spectroscopic values obtained with the  RAVESPAN software to check if the model is in agreement with the spectroscopic information (see Section~\ref{analysis}). We determined the spectroscopic luminosity ratio for  $\lambda$ = 6700 $\AA$ to be 0.934.  This luminosity ratio was then used to renormalize the disentangled spectra. Subsequently, the atmospheric analysis was repeated in order to obtain a better estimate of the effective temperatures and metallicities of the components (see Section~\ref{spec}).   
We derived effective temperatures of the components to be $T_1$=4860$\pm$185 K and $T_2$= 4700$\pm$180 K and metallicities of $[Fe/H]_1$=0.14$\pm$0.23 dex and $[Fe/H]_2$=0.39$\pm$0.28 dex.  We adopted $T_1$ as the new effective temperature of the primary component and repeated the fitting procedure using once again the DC subroutine of the WD code. After several iterations the best-fit model favoured a solution with the temperatures of the components nearly equal to each other. Therefore we decided to assume  a mean value of the effective temperature obtained from the atmospheric analysis as the temperature of the primary component. We set the $T_1$= 4780 K and repeated the fitting procedure. 
   
\begin{figure*}
\centering
        
        \includegraphics[width=\textwidth]{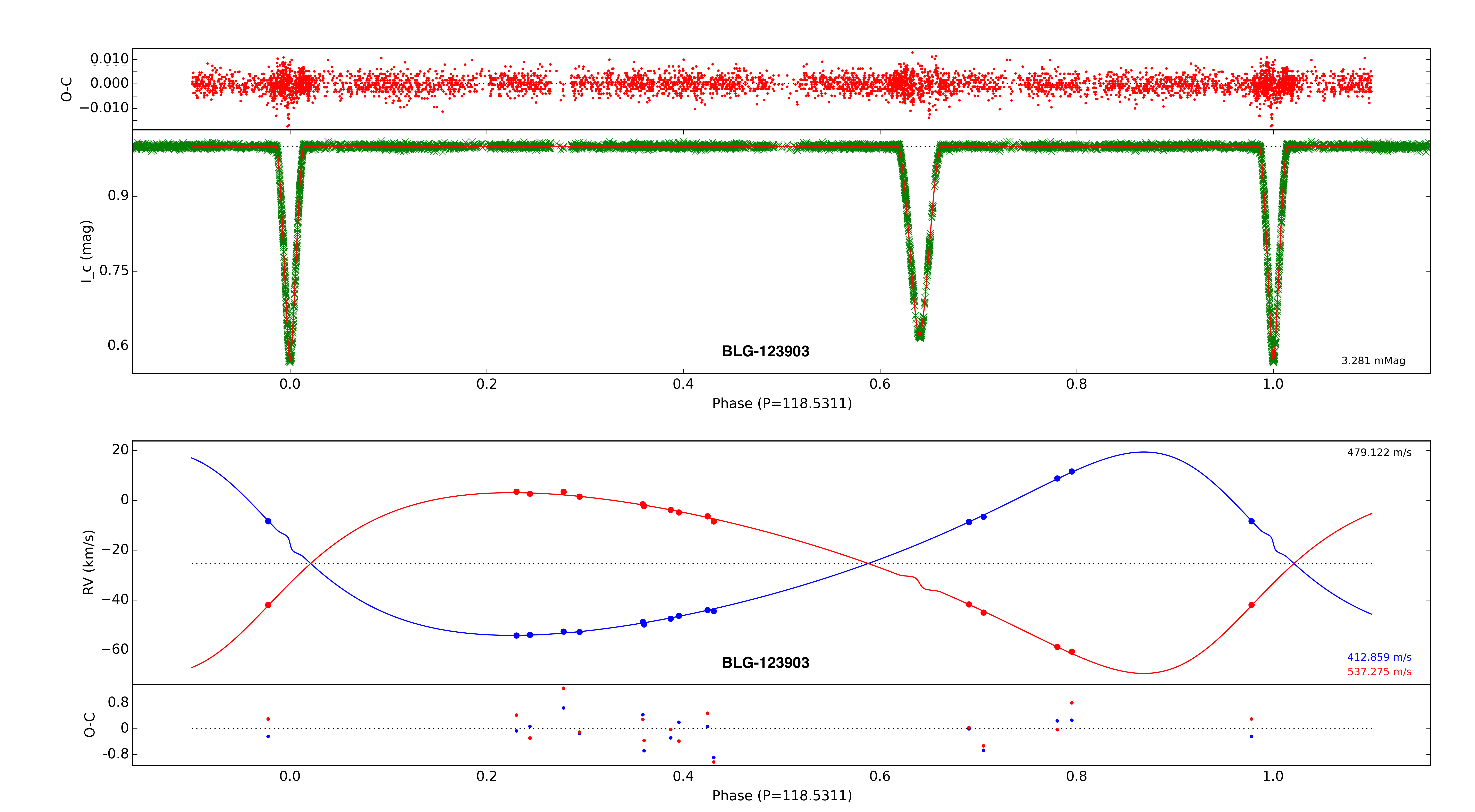}
    \caption{The $I$-band light curve and radial-velocity solutions from the WD code for OGLE-BLG-ECL-123903. The residuals of the fits are listed at the right ends of the panels. }
                \centering
    \label{fig:blg28_wd}
\end{figure*}

   The $I$-band light curve and radial-velocity solution from the WD code is presented in Fig.~\ref{fig:blg28_wd} and the derived parameters are presented in Table~\ref{table:results1}.
   
\subsubsection{BLG-296596}

The orbital period and the moment of primary minimum were adopted from the OGLE catalogue: $P$=155.638 d and $T_0$=2454244.3. The orbital period was later adjusted along with the phase-shift parameter (PSHIFT). We measured the average out-of-eclipse magnitudes using all the observational data outside the minima to be $V$ = 15.483 mag and $I$ = 13.074 mag. As before, we refer to the primary component as the one being eclipsed during the deeper, primary minimum.

We used the WD code to perform the modelling of the system. In the case of BLG-296596 we used only the $I$-band light curve, fit simultaneously with the radial velocity curve. The $V$-band OGLE photometry consists of only a few observational points and these were used to derive the out-of-eclipse magnitude. The same  parameters were adjusted in the fit
as for BLG-123903, with the exception of the luminosity ratio of the components, for
which only the $I$-band data were used.

We set the initial value of the temperature of the primary component to $T_1$=4500~K. It was our starting point to perform the iteration with the DC subroutine of the WD code in order to find the best-fit model of the system. During this procedure all of the free parameters were adjusted at the same time. 

At the end of the fitting procedure we also checked for the presence of  third light ($I_3$) to determine its impact on the solution. As for the case of BLG-123903, the third light
corrections were invariably negative, which suggested an unphysical solution, and therefore we adopted $I_3$=0 in the final solution. 

 The luminosity ratios from the solution were used to renormalize the disentangled spectra. Firstly, we compared them with the spectroscopic luminosity ratios obtained with the use of RaveSpan code. We determined the spectroscopic luminosity ratio for the $\lambda$ = 6700 $\AA$ to be 1.85. This value for the spectroscopic luminosity ratio is slightly lower than the value from the photometry solution for the same wavelength range. We readjusted the potentials of the components ($\Omega_1$ and $\Omega_2$) and fixed the potential parameter of the first component in order to get  consistent spectroscopic and photometric solutions. We repeated the fitting procedure, this time with one less free parameter. The luminosity ratios were then used to renormalize the disentangled spectra that were used to perform the atmospheric analysis enabling a better estimation of the effective temperatures of the components. We measured the temperatures to be $T_1$=4215$\pm$135 and $T_2$=4050$\pm$120 and the metallicities to be $[Fe/H]_1$=-0.02$\pm$0.23 and $[Fe/H]_2$=-0.11$\pm$0.19 for the primary and secondary component, respectively (see Table~\ref{tab.atmo2}). We then adopted the $T_1$ as a new effective temperature of the primary component and ran the DC subroutine of the WD code again in order to get a new best-fit model. 

\begin{figure*}
\centering
        \includegraphics[width=\textwidth]{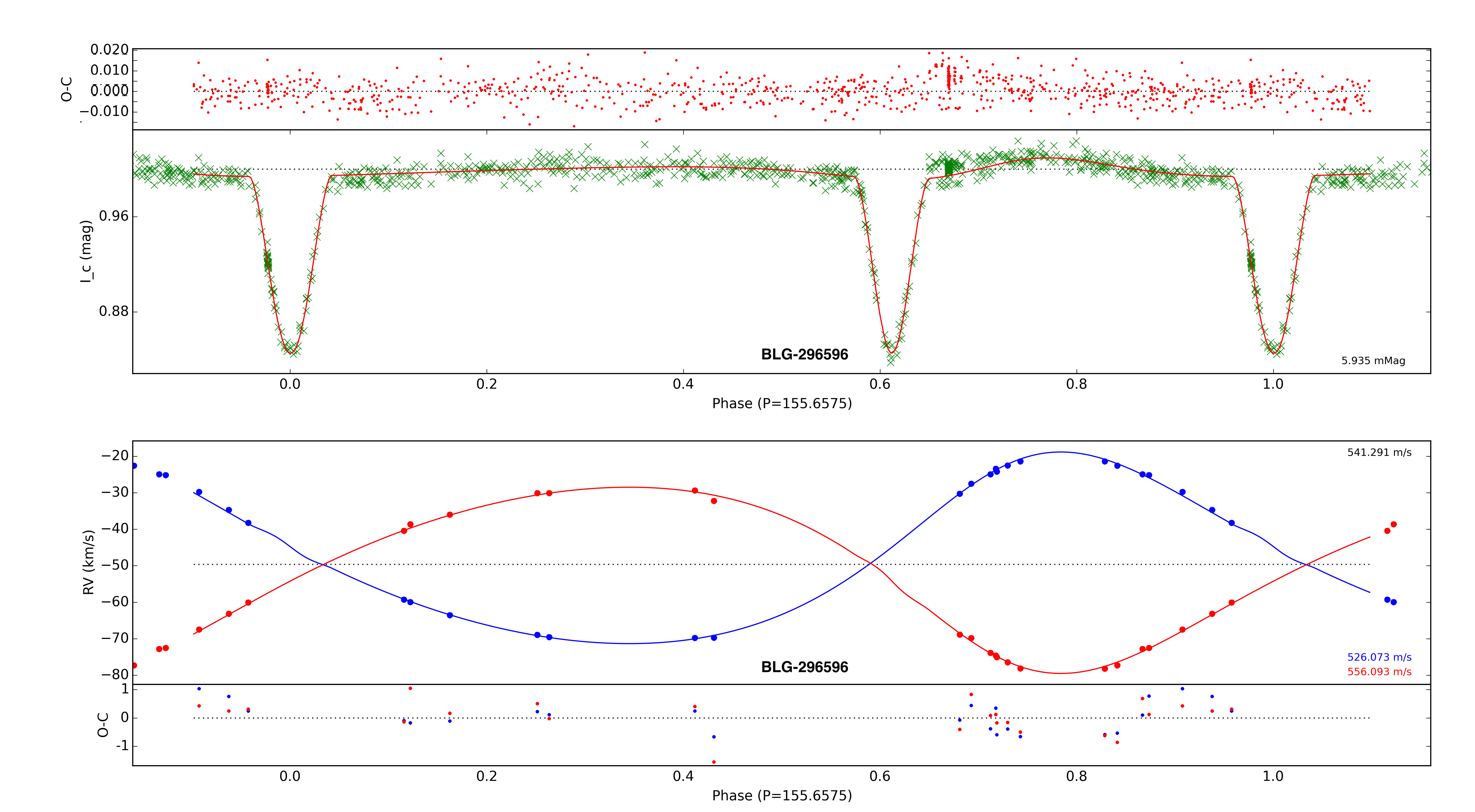}
    \caption{The $I$-band light curve and radial velocity solution from the WD code for OGLE-BLG-ECL-296596. The residuals of the fits are listed at the right ends of the panels. }
                \centering
    \label{fig:blg45_wd}
\end{figure*}

The $I$-band light curve and the radial-velocity solution from the WD code are presented in Fig.~\ref{fig:blg45_wd} and the derived parameters are presented in Table~\ref{table:results2}.

\begin{table} 
\caption{Photometric and orbital parameters obtained with the Wilson--Devinney code for the BLG-123903 system.}
 \begin{tabular}{p{0.61\linewidth}p{0.30\linewidth}}
\hline
Parameter & WD result \\ \hline \hline
Orbital inclination $i$ (deg) & 88.796 $\pm$ 0.007 \\
Orbital eccentricity $e$ & 0.3091 $\pm$ 0.0005\\
Sec. temperature $T_2$ (K) &  4786 $\pm$ 13\\
Fractional radius $r_1$ & 0.0586 $\pm$ 0.0002 \\
Fractional radius $r_2$ & 0.0556 $\pm$ 0.0003\\
$(r_1+r_2)$ & 0.1142 $\pm$ 0.0004 \\
$k = r_2/r_1$ & 0.9488 $\pm$ 0.0061\\
Observed period $P_{obs}$ (day) & 118.5311 $\pm$ 0.0015\\
$\omega$ (deg) & 45.533 $\pm$ 0.086\\
$\Omega_1$ & 18.568 $\pm$ 0.058\\
$\Omega_2$ & 19.683 $\pm$ 0.098\\
$(L2/L1)_V$ & 0.9101 $\pm$ 0.006\\
$(L2/L1)_I$ & 0.9084 $\pm$  0.006\\
$(L2/L1)_J$ & 0.9061\\
$(L2/L1)_K$ & 0.9053\\
$T_0$ (JD-2450000) & 7118.31  \\
Semimajor axis $a$ (R$_\odot$) & 162.81 $\pm$ 0.62 \\
Systemic velocity $\gamma$ (km s$^{-1}$) & -25.38 $\pm$ 0.05\\
Velocity semi-amplitude $K_1$(km s$^{-1}$) & 36.78 $\pm$ 0.17 \\
Velocity semi-amplitude $K_2$(km s$^{-1}$) &  36.25$\pm$ 0.22\\
Mass ratio $q$ &  1.015$\pm$  0.008\\ 
RV $rms_1$  (km s$^{-1}$) &0.412\\
RV $rms_2$  (km s$^{-1}$) &0.537\\ \hline
 \end{tabular}
 \label{table:results1}
\end{table}
 
\begin{table} 
\caption{Photometric and orbital parameters obtained with the Wilson--Devinney code for the BLG-296596 system.}
 \begin{tabular}{p{0.61\linewidth}p{0.30\linewidth}}
\hline
Parameter & WD result \\ \hline \hline
Orbital inclination $i$ (deg) & 79.86 $\pm$ 0.13 \\
Orbital eccentricity $e$ & 0.1964 $\pm$ 0.0013\\
Sec. temperature $T_2$ (K) &   3968$\pm$ 11\\
Fractional radius $r_1$ & 0.1137 $\pm$ 0.0017 \\
Fractional radius $r_2$ &  0.1876$\pm$ 0.0019\\
$(r_1+r_2)$ & 0.3013 $\pm$ 0.0025\\
$k = r_2/r_1$ & 1.6499 $\pm$ 0.0298\\
Observed period $P_{obs}$ (day) & 155.6576 $\pm$ 0.0043\\
$\omega$ (deg) & 331.33 $\pm$ 0.73\\
$\Omega_1$ (fixed) & 10.0844\\
$\Omega_2$ & 6.736 $\pm$ 0.049\\
$(L2/L1)_V$ &  1.6753 \\
$(L2/L1)_I$ & 1.9525$\pm$ 0.0587 \\
$(L2/L1)_J$ & 2.2375\\
$(L2/L1)_K$ & 2.5040\\
$T_0$ (JD-2450000) & 4244.34  \\
Semimajor axis $a$ (R$_\odot$) & 158.84 $\pm$ 0.66 \\
Systemic velocity $\gamma$ (km s$^{-1}$) & -49.608 $\pm$ 0.079\\
Velocity semi-amplitude $K_1$(km s$^{-1}$) & 26.27 $\pm$ 0.14\\
Velocity semi-amplitude $K_2$(km s$^{-1}$) & 25.55 $\pm$ 0.16\\
Mass ratio $q$ & 1.029 $\pm$  0.008\\ 
RV $rms_1$  (km s$^{-1}$) &0.594\\
RV $rms_2$  (km s$^{-1}$) &0.574\\ \hline
 \end{tabular}
 \label{table:results2}
\end{table}

  \begin{table}
  \caption{Physical properties of the BLG-123903 system.}
\centering
  \begin{tabular}{p{0.3\linewidth}p{0.30\linewidth}p{0.24\linewidth}}
  \hline
  Property & The Primary & The Secondary \\ \hline \hline
  Spectral type & K2 III & K2 III\\
  $V^a$ (mag) & 17.607 & 17.709 \\
  $I^a$ (mag) & 14.785 & 14.889 \\
  $J^a$ (mag) & 12.697 & 12.804 \\
  $K^a$ (mag) & 11.529 & 11.637 \\
  $V\!-\!I$ (mag) & 2.822 & 2.820 \\
  $V\!-\!K$ (mag) & 6.077 & 6.072 \\
  $J\!-\!K$ (mag) & 1.168 & 1.167 \\
  Radius ($R_{\odot}$) & 9.540$\pm$ 0.049  & 9.052$\pm$ 0.060 \\
  Mass ($M_{\odot}$) & 2.045 $\pm$  0.027 & 2.074 $\pm$ 0.023\\
  log $g$  & 2.790 $\pm$ 0.011 & 2.841 $\pm$ 0.012  \\
  $T_{\rm eff}$ (K) & 4780$^b$ $\pm$ 131 & 4786$^c$ $\pm$ 180 \\
  $v$ sin $i$ (km s$^{-1}$) & 8.86 $\pm$ 1.51 & 8.86 $\pm$ 1.44  \\
  Luminosity ($L_{\odot}$) & 43 $\pm$ 7 & 39 $\pm$ 6\\
  $M_{\rm bol}$ (mag) & 0.684 & 0.789\\
  $M_{\rm v}$ (mag) & 1.028 & 1.131 \\
  $[$Fe/H$]^b$ & 0.14 $\pm$ 0.23& 0.39 $\pm$ 0.28\\ \hline
  $E(B\!-\!V)$ & 1.329 $\pm$ 0.114 \\
  Distance (pc) & 2953 $\pm$  59.1(stat.) & $\pm$ 70.2 (syst.) \\ 
 \hline
 $^{a -\rm  observed}$ 
 $^{b- \rm atmospheric\: analysis}$
 $^{c -\rm WD\: solution}$ 
  \end{tabular}
  \centering
  \label{tab:results_final1}
\end{table}

 \begin{table}
  \caption{Physical properties of the BLG-296596 system.}
\centering
  \begin{tabular}{p{0.28\linewidth}p{0.35\linewidth}p{0.24\linewidth}}
  \hline
  Property & The Primary & The Secondary \\ \hline \hline
  Spectral type & K6 III & K8 III\\
  $V^a$ (mag) & 16.551 & 15.991 \\
  $I^a$ (mag) & 14.249 & 13.523 \\
  $J^a$ (mag) & 12.623 & 11.749 \\
  $K^a$ (mag) & 11.518 & 10.521 \\
  $V\!-\!I$ (mag) & 2.302 & 2.468 \\
  $V\!-\!K$ (mag) & 5.034 & 5.470 \\
  $J\!-\!K$ (mag) & 1.105 & 1.228 \\
  Radius ($R_{\odot}$) & 18.06$\pm$ 0.28 & 29.80 $\pm$ 0.33 \\
  Mass ($M_{\odot}$) & 1.093 $\pm$  0.015& 1.125 $\pm$ 0.014 \\
  log $g$  & 1.963 $\pm$ 0.013 & 1.541 $\pm$ 0.011  \\
  $T_{\rm eff}$ (K) & 4215$^b$ $\pm$  135& 3968$^c$ $\pm$ 120 \\
  $v$ sin $i$ (km s$^{-1}$) & 8.76 $\pm$ 1.29 & 18.72 $\pm$  1.31 \\
  Luminosity ($L_{\odot}$) & 92 $\pm$ 14&198  $\pm$ 27\\
  $M_{\rm bol}$ (mag) & -0.193 & -1.107\\
  $M_{\rm v}$ (mag) & 0.534 &  -0.026\\
  $[$Fe/H$]^b$ & -0.02 $\pm$ 0.23& -0.11 $\pm$ 0.19\\ \hline
  $E(B\!-\!V)$ & 0.731$\pm$ 0.117 \\
  Distance (pc) & 5682.8 $\pm$  74.2(stat.) & $\pm$  142.3  (syst.) \\ 
 \hline
 $^{a - \rm observed}$ 
 $^{b- \rm atmospheric\: analysis}$
 $^{c -\rm WD\: solution}$ 
  \end{tabular}
  \centering
  \label{tab:results_final2}
\end{table}

\subsection{Absolute dimensions}
The final values  of the physical parameters of the components of both systems are presented in Tables~\ref{tab:results_final1} and~\ref{tab:results_final2}. 

We derived masses of the components based on the following equations:

\begin{equation}
M_1[M_\odot] = 1.34068 \cdot 10^{-2} \frac{1}{1+q}\frac{a^3[R\odot]}{P^2[d]}
,\end{equation}

\begin{equation}
M_2 [M_\odot] = M_1 \cdot q
,\end{equation}

where $q$ is the mass ratio, $a$ is the semi major axis, and $P$ is the orbital period.  
The individual observed magnitudes and the colour relations of the components of both systems presented in Tables ~\ref{tab:results_final1} and ~\ref{tab:results_final2} were derived based on the equations:

\begin{equation}
\label{V1}
V_1=V - 2.5 \log{ \frac{1}{1+\frac{L_2}{L_1}}}
,\end{equation}

\begin{equation}
\label{V2}
V_2= V - 2.5\log{ \frac{\frac{L_2}{L_1}}{1+\frac{L_2}{L_1}}}
,\end{equation}
where $V_1$ and $V_2$ are the magnitudes of the primary and secondary components and $\frac{L_2}{L_1}$ is the luminosity ratio in a given filter. The magnitudes for $I$, $J$, $K$ band filters were determined in a similar fashion. 
We converted the $V$-band magnitudes to bolometric magnitudes using the bolometric corrections from \cite{alo99}.

\section{Evolutionary status}
 \label{sec_evolution}

We examined the evolutionary status of BLG-123903 and BLG-296596  using published \textsc{parsec} isochrones \citep{bres12} and a small grid of \textsc{mesa} \citep{Paxton18} tracks computed specifically for the two systems.

Isochrones were fitted as described in \cite{suchomska15}. For a given metallicity, the isochrone (age) was selected to minimise the $\chi^2$ function including luminosities, effective temperatures, and radii of the two components. Model values were calculated at mass points corresponding to the masses of the components of each 
binary system.

The evolutionary calculations were performed using publicly available and open source code \textsc{mesa} \citep{Paxton11,Paxton13,Paxton15,Paxton18}, version 10108. All models were calculated with the OPAL opacity tables \citep{Iglesias96} and the opacity tables of \cite{Ferguson05} for the lower temperatures. The mixture of the heavier elements is adopted from \cite{Asplund09}. The adopted nuclear reaction rates are taken from the JINA REACLIB database \citep{Cyburt10,Rauscher00}. The outer boundary conditions are based on the photospheric tables constructed with the \textsc{phoenix} code \citep{Hauschildt99a,Hauschildt99b} and supplemented with the tables of \cite{Castelli03} for higher temperatures. The borders of convective zones are determined by Ledoux criterion for the convective instability and the overshooting from convective zones is treated within the exponential model of \cite{Herwig00}. Metallicity, $\mathrm{[Fe/H]}$, the efficiency of overshooting from the hydrogen core, $f_\mathrm{H}$, and the efficiency of inward overshooting from the outer convective zone, $f_\mathrm{env}$, were treated as variables. Other parameters were fixed, including the masses of the components. The adopted value for the mixing length parameter is $\alpha_\mathrm{MLT} = 1.80$, based on the calibration of \cite{Ostrowski18}. More details on the modelling, underlying physics, and fitting procedure can be found in \cite{Ostrowski18}.

\subsection{BLG-123903} \label{sec_evolution_123903}

\begin{figure}
\centering
\includegraphics[width=\columnwidth]{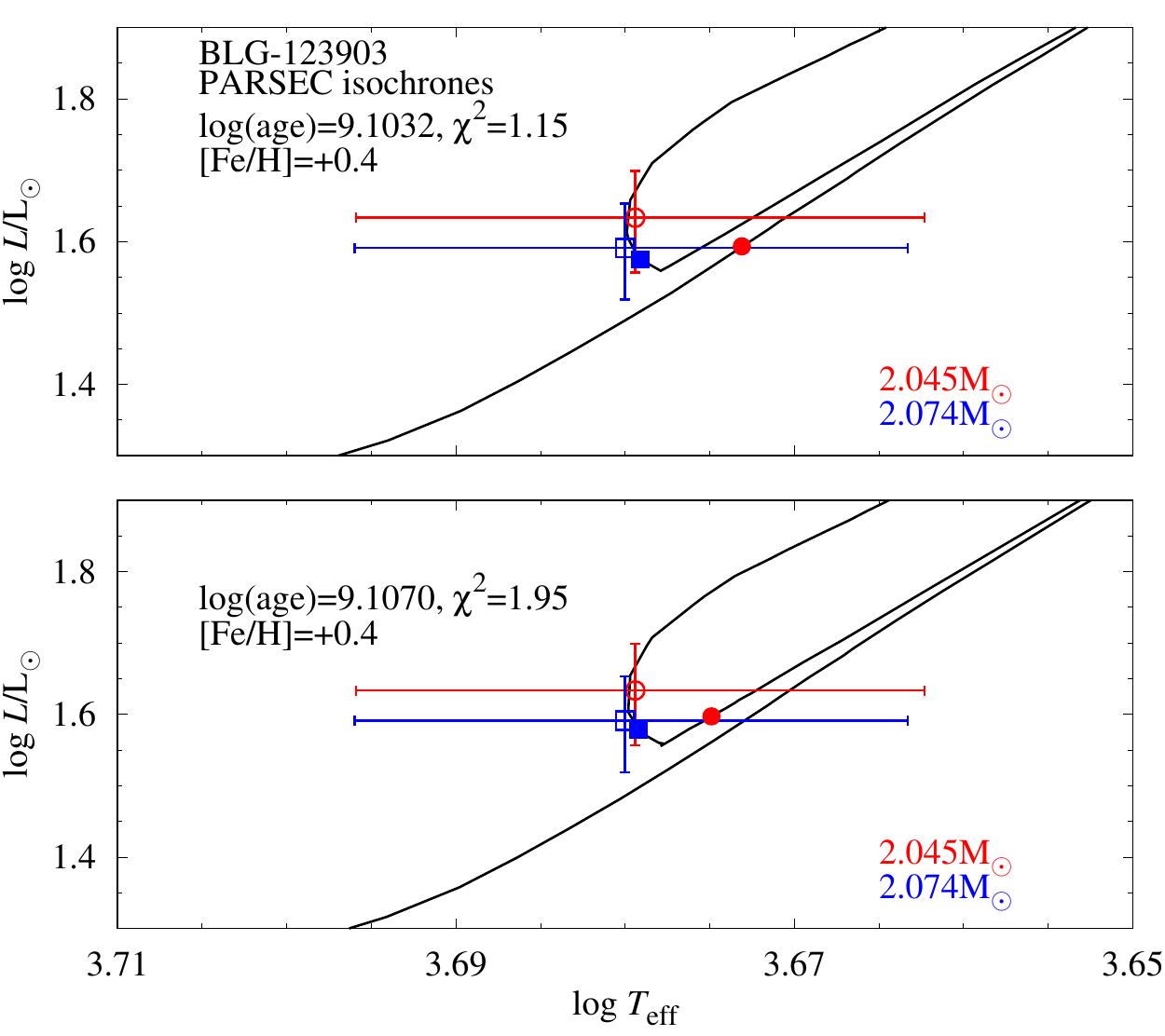}
\caption{\textsc{parsec} isochrones for BLG-123903 and ${\rm [Fe/H]}=+0.4$. The location of the primary and secondary components is marked with circles and squares, respectively. Filled symbols refer to the fitted isochrones and the open ones to our measurements. Two solutions of similar quality are shown in the top and bottom panels.}
\label{fig:hmhr}
\end{figure}

Two solutions of similar $\chi^2$ value based on \textsc{parsec} isochrones are presented in Fig.~\ref{fig:hmhr}. The best fit is obtained for isochrones with high metallicity, ${\rm [Fe/H]}=+0.4$, which agrees well with the metallicity determined from the spectroscopic analysis. The location of the two components in the HR diagram is matched within $1\,\sigma$ error boxes and the age is $\approx 1.3$\,Gyr in both solutions. In the top panel of Fig.~\ref{fig:hmhr} the more massive component is steadily burning helium in its core, while the less massive component is less evolved and is climbing the RGB. The solution plotted in the bottom panel of Fig.~\ref{fig:hmhr} is more probable: both components are steadily burning helium in their cores. This solution is consistent with evolutionary modelling discussed below. Results are qualitatively the same regardless of whether we use isochrones with or without mass loss included.

\begin{figure*}
\begin{center}
\includegraphics[clip,width=86mm,angle=0]{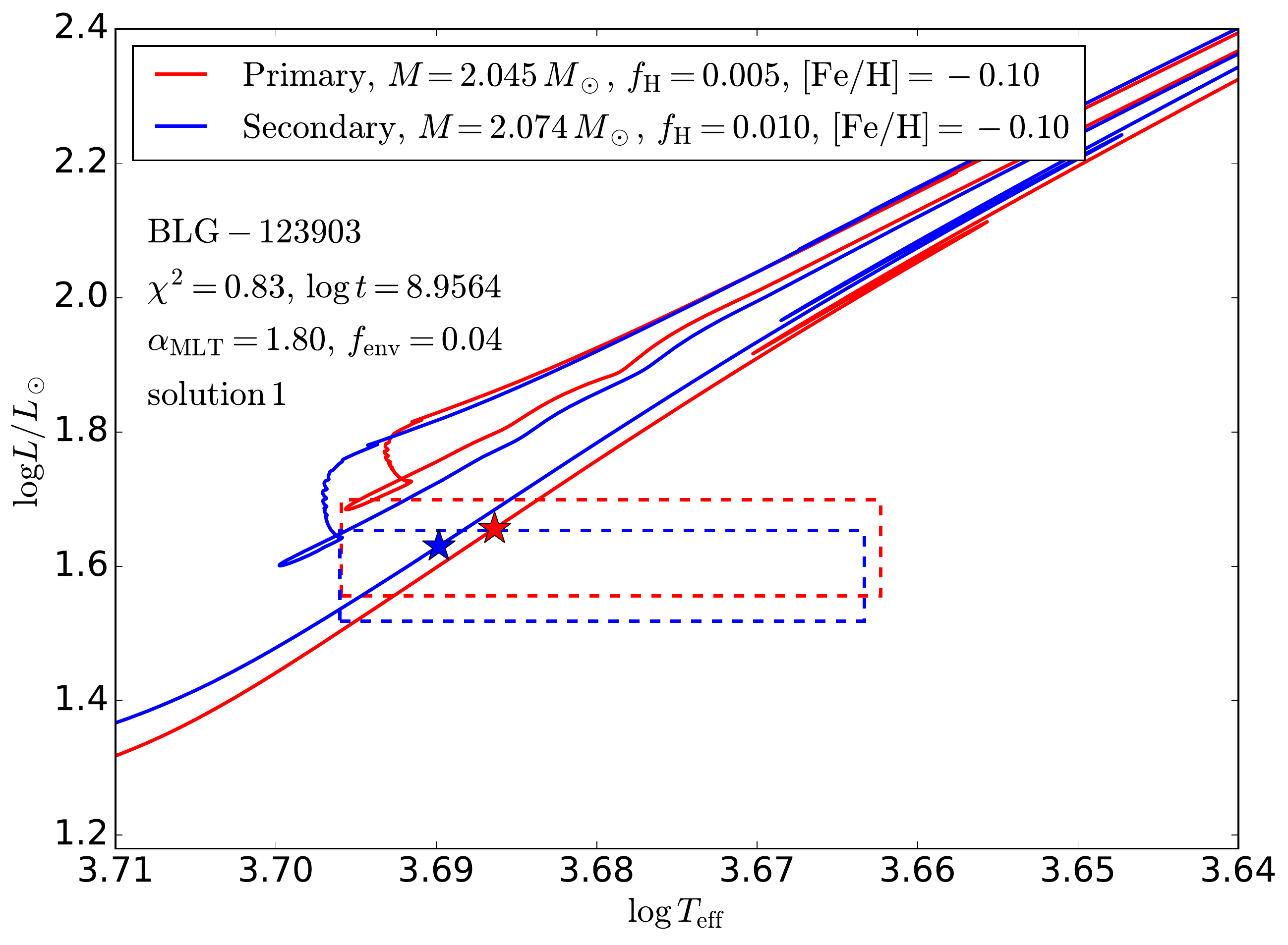}
\includegraphics[clip,width=86mm,angle=0]{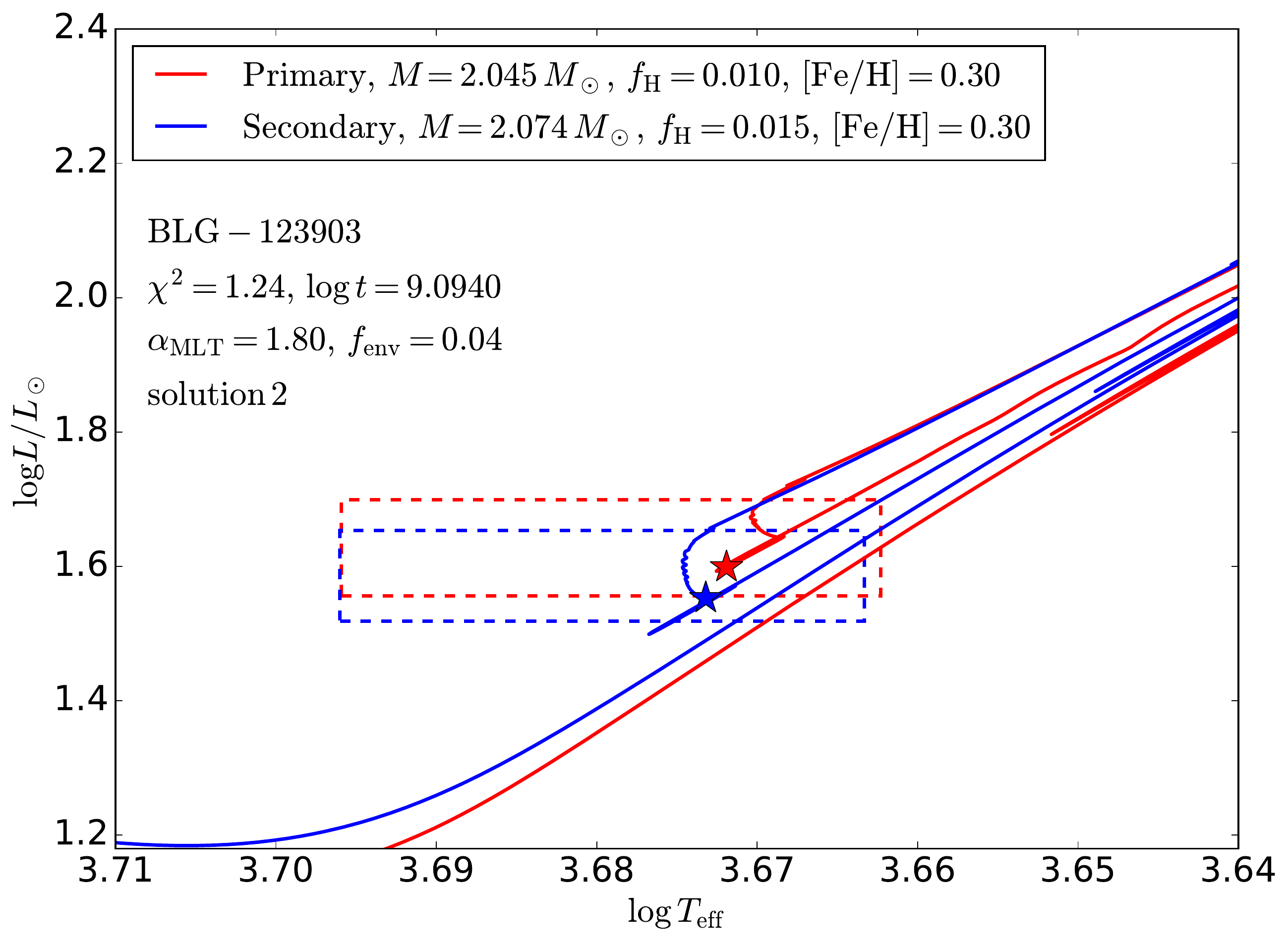}
\caption{HR diagrams with sections of \textsc{mesa} evolutionary tracks around two solutions for the system BLG-123903. Red lines are used for the primary component and blue lines for the secondary component. Red and blue dashed rectangles show the $1\sigma$ error boxes for primary and secondary component, respectively, and red and blue asterisks mark the locations of the models fitted for the primary and the secondary, respectively.}
\label{fig_blg_123903_hr}
\end{center}
\end{figure*}

Evolutionary modelling with \textsc{mesa} yields two solutions of very good quality for BLG-123903. In the left panel in Fig.\,\ref{fig_blg_123903_hr}, we show the first solution, for which both stars are ascending the RGB (central helium abundances are $Y_\mathrm{c,\,p} = Y_\mathrm{c,\,s} = 0.9891$). Red lines depict the primary component and blue lines  the secondary component. Red and blue dashed rectangles show the $1\sigma$ error boxes for the primary and secondary component, respectively, and red and blue asterisks mark the locations of the models fitted for the primary and the secondary, respectively. The derived metallicity, $\mathrm{[Fe/H]} = -0.10$, is more than $1\sigma$ lower than the luminosity-weighted average metallicity obtained for the system from observations, $\mathrm{[Fe/H]_\mathrm{avg}} =+0.25 \pm 0.25$. The derived age of the system is $\log t = 8.9564$~yr. The efficiencies of convective overshooting from the hydrogen core are small with $f_\mathrm{H,\,p} = 0.005$ and $f_\mathrm{H,\,s} = 0.010$ for the primary and the secondary, respectively, and the inward overshooting from the outer convective zone has an efficiency of $f_\mathrm{env} = 0.04$. In this solution, both components are climbing the RGB, but the more massive secondary component is less luminous. According to the standard theory of evolution, the more massive component should evolve faster than the less massive star, but this is under the implicit assumption that both stars are otherwise identical, that is, that they have the same chemical composition, undergo the same mixing processes, and so on. In case of the considered system and the adopted solutions, these assumptions may not be fulfilled. The stars have the same age and metallicity, but the efficiency of overshooting from the hydrogen core during the main-sequence evolution is higher for the more massive component. The difference in the masses between the components is very small and hence the difference in the mixing efficiency leads to the slower rate of evolution of the slightly more massive star. The less massive primary component (M = 2.045 M$_\odot$) spent 858 Myr during core hydrogen burning phase, whereas the more massive secondary component (M = 2.074 M$_\odot$) spent 876 Myr due to the more efficient mixing. Such differences in the evolutionary time on a very long MS phase influence the later, shorter stages and explain the seemingly strange behaviour on the RGB. Stars with similar masses may have different overshooting efficiencies because in the framework of the paper we do not explicitly include the effects of rotation in the calculated models. The different overshooting efficiencies may represent the additional mixing related to the omitted rotationally induced mixing.

The second \textsc{mesa} solution (Fig.\,\ref{fig_blg_123903_hr}, right panel) places both stars in the core-helium burning phase. The more massive secondary component commences stable helium burning at the red clump ($Y_\mathrm{c,\,s} = 0.9095$) and the less massive primary is just beginning this phase of evolution ($Y_\mathrm{c,\,p} = 0.9514$). The derived age is $\log t = 9.0940$. The metallicity, $\mathrm{[Fe/H]} = +0.30$, is higher than in the previous solution and is in  better agreement with the observations, shifting the evolutionary tracks toward lower effective temperatures. The efficiencies of core overshooting are also slightly higher: $f_\mathrm{H,\,p} = 0.010$ and $f_\mathrm{H,\,s} = 0.015$, whereas the overshooting from the outer convective zone during RGB evolution has the same efficiency as in the previous solution.

\subsection{BLG-296596} \label{sec_evolution_296596}

\begin{figure}
\centering
\includegraphics[width=\columnwidth]{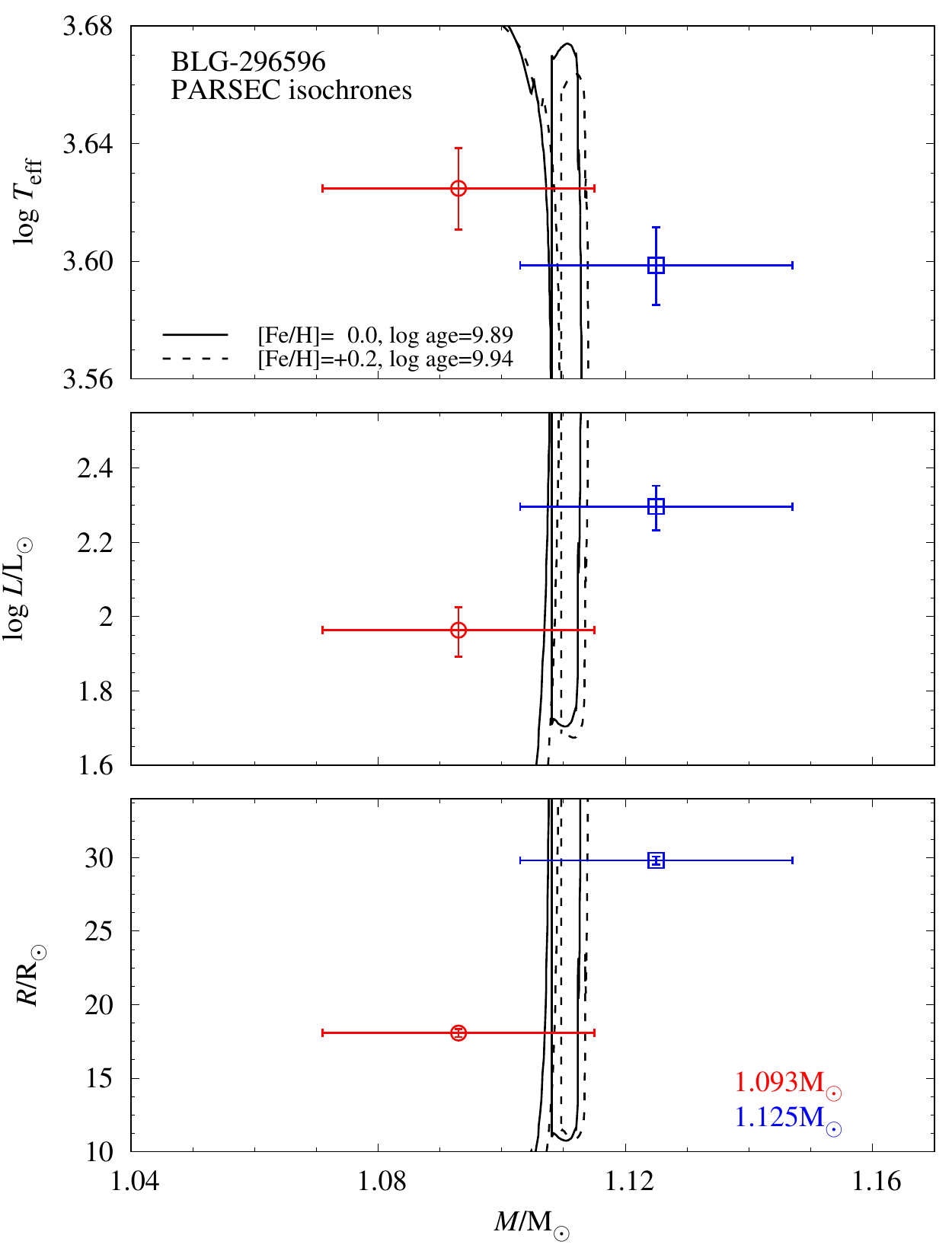}
\caption{\textsc{parsec} isochrones for BLG-296596. In the consecutive panels, effective temperature, luminosity, and radius are plotted as a function of mass. Solid and dashed lines correspond to ${\rm [Fe/H]}=0.0$, log age = 9.89 and  ${\rm [Fe/H]}=+0.2$, log age = 9.94, respectively.}
\label{fig:lmiso}
\end{figure}

\begin{figure}
\centering
\includegraphics[width=\columnwidth]{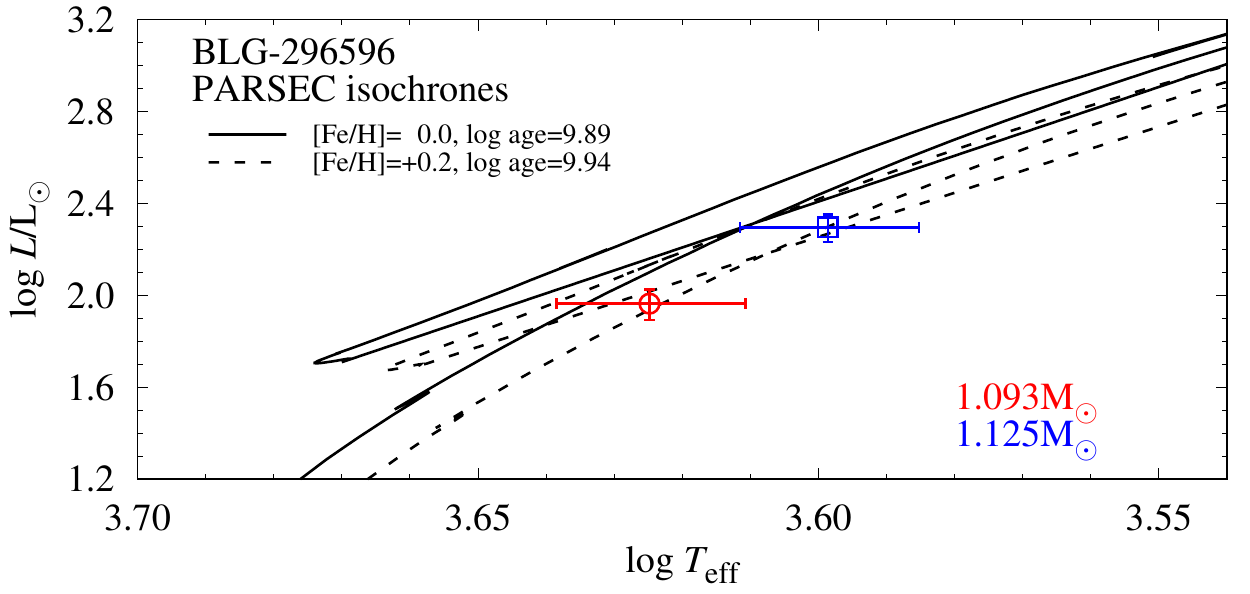}
\caption{\textsc{parsec} isochrones for BLG-296596 plotted in the HR diagram.}
\label{fig:lmhr}
\end{figure}

We were not able to find solutions for the BLG-296596 using \textsc{mesa} models and the measured  masses of the components,  1.093\,${\rm M}_\odot$ and 1.125\,${\rm M}_\odot$; neither  could we select a single \textsc{parsec} isochrone encompassing the exact values of the two masses at the post-main sequence evolution.
Theoretical models indicate that low-mass, evolved, post-main sequence binaries should be composed of nearly equal-mass components. This is illustrated in Fig.~\ref{fig:lmiso} with the help of the \textsc{parsec} isochrones for ${\rm [Fe/H]}=0.0$, log age = 9.89 and  ${\rm [Fe/H]}=+0.2$, log age = 9.94. Effective temperature, luminosity, and radius are plotted as a function of mass. The larger the mass, the faster the evolution and the later the evolutionary stage along the isochrone. It is immediately visible that isochrones run almost vertically in the plots. The three isochrone sections from left to right correspond to the RGB ascent, descent after helium ignition at the tip of the RGB, and finally to the ascent along the AGB after helium gets depleted in the core. Although the plotted isochrones match both masses of BLG-296596 within $1\,\sigma$ errors, it is clear that an exact match of the two masses with a single isochrone is not possible.  In Fig.~\ref{fig:lmhr} we show the same isochrones in the HR diagram. The most metal-rich isochrone best matches the location of the two components. Since a  more massive component is more luminous, and the system has been detached during evolution, it is most probable that the two components of BLG-296596 increase their luminosity. As the evolutionary time-scale is longer for RGB ascent than for the AGB ascent, the former evolutionary stage is more likely.

We note that the isochrones predict nearly equal masses for the components of this low-mass post-main sequence binary system: the masses should be even closer than the mass ratio predicts ($q=1.029\pm 0.008$). Further observations and refinement of the parameters of this system are needed to assess the state of this system compared with evolutionary models. 

\begin{table*}
 \centering
  \caption{Error budget of the distance moduli of BLG-123903.}

 \resizebox{\textwidth}{!}{\begin{tabular}{@{}cccccccccc@{}}
  \hline
      Type of error & $(m - M)$ & $\sigma$A   & $\sigma$diBenedetto & $\sigma E(B-V)$  & $\sigma V$ & $\sigma K$  & $(L_2/L_1)_K$ &Combined Error  \\
    &(mag)&(mag)&(mag)&(mag)&(mag)&(mag)&(mag) &(mag) \\ \hline \hline
   \textbf{Statistical}&12.352 & 0.003& --& $0.036^1$& 0.001& 0.024& -&\textbf{0.043}\\
   \textbf{Systematic} & 12.352&--& 0.040 &--&\multicolumn{2}{c}{0.03}& 0.01&\textbf{0.051} \\
  \hline
  \multicolumn{3}{c}1 - combination of statistical and systematic error
  \end{tabular}}
  \centering
  \label{tab:error_1}
\end{table*}

 \begin{table*}
 \centering
  \caption{Error budget of the distance moduli of the BLG-296596.}

\resizebox{\textwidth}{!}{ \begin{tabular}{@{}cccccccccc@{}}
  \hline
      Type of error & $(m - M)$ & $\sigma$A   & $\sigma$diBenedetto & $\sigma E(B-V)$  & $\sigma V$ & $\sigma K$  & $(L_2/L_1)_K$ &Combined Error  \\
    &(mag)&(mag)&(mag)&(mag)&(mag)&(mag)&(mag) &(mag) \\ \hline \hline
   \textbf{Statistical}&  13.773&0.006 &--& $0.013^1$& 0.002&0.024 & -&\textbf{0.028}\\
   \textbf{Systematic} & 13.773&--& 0.044&--&\multicolumn{2}{c}{0.03}& 0.01&\textbf{0.054} \\
  \hline
  \multicolumn{3}{c}1 - combination of statistical and systematic error
  \end{tabular}}
  \centering
  \label{tab:error_blg}
\end{table*}

\subsection{Distances to the BLG-123903 and BLG-296596}
\label{distance}
To determine the distances to the analysed systems we used the surface brightness--color relation (SBCR) measured for Galactic late-type giant stars by \cite{ben05}. We followed the prescription presented in \cite{pietrzyn2009}. 
The resulting distance to BLG-123903 is $d$ = 2953 $\pm$ 59.1(stat.) $\pm$ 70.2(syst.) pc and for the BLG-296596  is  $d$ = 5682.8 $\pm$ 74.2(stat.) $\pm$ 142.3(syst.). 

The main statistical uncertainty comes from the uncertainties in the measured infrared photometry and reddening. Therefore, there is significant room for improvement in the derived distance with more precise infrared photometry. The main contribution to the systematic error comes from the SBCR calibration. The contributions to the statistical and systematical error for  distance measurements for both systems is presented in Tables \ref{tab:error_1}   and \ref{tab:error_blg} for BLG-123903 and BLG-296596,  respectively.

According to \cite{gaia18}, the parallaxes for our investigated systems in the Gaia Data Release 2 (DR2) were measured to be $\varpi$ = 0.3173 $\pm$ 0.0734 $mas$ and $\varpi$ = 0.2165 $\pm$ 0.0492 $mas$ for  BLG-123903 and BLG-296596, respectively. Our measured distances correspond to parallaxes of $\varpi_{phot}$ = 0.3386 and $\varpi_{phot}$ = 0.1759 $mas$ for  BLG-123903 and BLG-296596, respectively.  The measurements are in agreement within the margin of error. For  BLG-296596, at a larger distance, the divergence in the measurements is much higher, almost at the margin of error. 

We also compared our distance determinations with those presented by \cite{bailer18} based on  parallaxes in Gaia DR2. According to the authors,  reliable distances cannot be obtained by inverting the parallax for the majority of the stars in  Gaia DR2. They used  an inference procedure to account for the non-linearity of the transformation and the asymmetry of the resulting probability distribution by adopting a weak distance prior that varies smoothly as a function of Galactic longitude and latitude according to a Galaxy model. \cite{bailer18} estimate the distance to BLG-123903 to be d= 2988$_{-602}^{+980}$ pc and to BLG-296596 to be d= 4073$_{-732}^{+1100}$ pc. For  BLG-123903 the estimated distance is in perfect agreement with our determination within the margin of error. For BLG-296596, the distance presented by \cite{bailer18} is lower than the value presented in this paper by 1.6~kpc.  Therefore we note that the weak distance prior used in this latter study is not efficient when it comes to this particular system. In this case the direct comparison of the Gaia parallaxes with the photometric parallaxes obtained from our distance determinations are more accurate. The distance determinations presented in this section provide an independent way of testing Gaia distance and parallax measurements.

\section{Summary And Conclusions}

We have measured the physical and orbital parameters for two well-detached eclipsing binary systems OGLE-BLG-ECL-123903 and OGLE-BLG-ECL-296596. The accuracy of our estimations was between 1 and 2 \%.  We also measured the distance to the systems and derived  $d$ = 2.95 $\pm$ 0.06 (statistical error) $\pm$ 0.07 (systematic error) kpc for  BLG-123903, and $d$ = 5.68 $\pm$ 0.07 (stat.) $\pm$ 0.14 (syst.) for BLG-296596. The accuracy of the distance determinations is  $\sim$3\%, which is comparable but somewhat less accurate than the distances to the Large/Small Magellanic Cloud binaries obtained using the same method. This is caused mainly by the much higher interstellar extinction, as well as uncertainties in the infrared photometry for this sample of binaries.

The used method of distance determinations serves also as an independent way
 of testing the distance and parallax determinations provided by the Gaia mission. 
Our results, based on observations of evolved component stars in detached eclipsing binary stars, are important for testing  stellar evolution theory. So far only a few such systems with well-determined parameters have been studied \citep{pie13, hel15}. Calculations of evolutionary status of evolved stars are sensitive to many parameters, and therefore the observation and analysis of additional systems is much needed.

\begin{acknowledgements}
We would like to thank the staffs of the Las Campanas and ESO La Silla Observatories for their support during the observations. 
We gratefully acknowledge financial support for this work from the Polish National Science Centre grants OPUS DEC-2013/09/B/ST9/01551. We also acknowledge support from the IdP II 2015 0002 64 grant of the Polish Ministry of Science and Higher Education. WG, GP and S.V gratefully acknowledge financial support for this work
from the Chilean Centro de Excelencia en Astrofisica y Tecnologias
Afines (CATA) BASAL grant AFB-170002. WG, DG and MG also acknowledge
support from the Millennium Institute of Astrophysics (MAS) of the Iniciativa Cientifica Milenio del Ministerio de Economia, Fomento y Turismo de Chile, project IC120009. K.S. acknowledges the financial support from the National Science Centre under the grant ETIUDA 2016/20/T/ST9/00174. S.V gratefully acknowledges the support provided by Fondecyt reg. n. 1170518. This work is based on observations collected at the European Organisation for Astronomical Research in the Southern Hemisphere under ESO programmes: 095.D-0026(A), 092.D-0363(A). We are pleased to thank the ESO,  Chilean, and Carnegie Time Assignment Committees for their generous support of this program.
\end{acknowledgements}






\label{lastpage}
\end{document}